\pdfoutput=1
\documentclass[%
reprint,
superscriptaddress,
 amsmath,amssymb, 
 aps,
 prl,
floatfix,
]{revtex4-2}

\usepackage{graphicx}
\usepackage{epsfig} 
\usepackage{dcolumn}
\usepackage{bm}
\usepackage{epstopdf}
\usepackage{multirow}
\usepackage{amsmath}
\usepackage{booktabs}
\usepackage{color}
\usepackage{gensymb}
\usepackage{kantlipsum} 
\usepackage{hyperref}
\usepackage{ulem}
\usepackage{braket}
\usepackage{physics}
\usepackage[utf8]{inputenc}
\usepackage[T1]{fontenc}
\usepackage{mathptmx}

\definecolor{darkviolet}{RGB}{148,0,211}

\begin{document}

\title{Observation of Bose-Einstein Condensation of Dipolar Molecules}

\preprint{APS/123-QED}


\author{Niccol\`{o} Bigagli}
\thanks{These authors contributed equally.}
\affiliation{Department of Physics, Columbia University, New York, New York 10027, USA}
\author{Weijun Yuan}
\thanks{These authors contributed equally.}
\affiliation{Department of Physics, Columbia University, New York, New York 10027, USA}
\author{Siwei Zhang}
\thanks{These authors contributed equally.}
\affiliation{Department of Physics, Columbia University, New York, New York 10027, USA}
\author{Boris Bulatovic}
\affiliation{Department of Physics, Columbia University, New York, New York 10027, USA}
\author{Tijs Karman}
\affiliation{Institute for Molecules and Materials, Radboud University, 6525 AJ Nijmegen, Netherlands}
\author{Ian Stevenson}
\affiliation{Department of Physics, Columbia University, New York, New York 10027, USA}
\author{Sebastian Will}\email{Corresponding author. Email: sebastian.will@columbia.edu}
\affiliation{Department of Physics, Columbia University, New York, New York 10027, USA}


\begin{abstract}

Ensembles of particles governed by quantum mechanical laws exhibit fascinating emergent behavior. Atomic quantum gases \cite{anderson1995observation, davis1995bose}, liquid helium \cite{kapitza1938viscosity, allen1938flow}, and electrons in quantum materials \cite{tsui1982two, bednorz1986possible, cao2018unconventional} all show distinct properties due to their composition and interactions. Quantum degenerate samples of bosonic dipolar molecules promise the realization of novel phases of matter with tunable dipolar interactions \cite{baranov2012condensed} and new avenues for quantum simulation~\cite{micheli2006toolbox} and quantum computation~\cite{demille2002quantum}. However, rapid losses \cite{ospelkaus2010quantum}, even when reduced through collisional shielding techniques \cite{valtolina2020dipolar, matsuda2020resonant, anderegg2021observation}, have so far prevented cooling to a Bose-Einstein condensate (BEC). In this work, we report on the realization of a BEC of dipolar molecules. By strongly suppressing two- and three-body losses via enhanced collisional shielding, we evaporatively cool sodium-cesium (NaCs) molecules to quantum degeneracy. The BEC reveals itself via a bimodal distribution and a phase-space-density exceeding one. BECs with a condensate fraction of 60(10) $\%$ and a temperature of 6(2)~nK are created and found to be stable with a lifetime close to 2 seconds. This work opens the door to the exploration of dipolar quantum matter in regimes that have been inaccessible so far, promising the creation of exotic dipolar droplets~\cite{schmidt2022self}, self-organized crystal phases~\cite{buchler2007strongly}, and dipolar spin liquids in optical lattices~\cite{yao2018quantum}. 
\end{abstract}

\maketitle

\section{Introduction}
The behavior of many-body quantum systems is dictated by the interactions between their constituents. Starting from weak contact interactions, Bose-Einstein condensates (BECs) of neutral atoms \cite{anderson1995observation,davis1995bose} were instrumental in gaining full control of atoms, in understanding the basic properties of superfluids, and in establishing quantum simulation as a viable experimental method \cite{zwerger2011bcs,gross2017quantum}. BECs of highly magnetic atoms enabled the study of systems with weak long-range dipole-dipole interactions \cite{chomaz2022dipolar} and the realization of rich phases of matter such as quantum ferrofluids \cite{lahaye2007strong} and droplets \cite{kadau2016observing, chomaz2016quantum}. Even stronger interactions are at the core of the intriguing properties of liquid helium \cite{wilks1967properties} or electron gases in solid state systems \cite{bardeen1957theory, laughlin1983anomalous}.

Quantum gases of ground state dipolar molecules have been proposed as a clean and controlled system in which long-range interactions can be tuned from the weakly interacting to the strongly interacting regime. In the limit of weak interactions, bosonic molecules will form a Bose-Einstein condensate, much like dilute atomic clouds. Once the strength of interactions is increased, theoretical predictions abound \cite{baranov2012condensed} and include the realization of strongly correlated phases of matter \cite{schmidt2022self}, from supersolids \cite{pollet2010supersolid} to dipolar crystals \cite{buchler2007strongly} and Mott insulators with fractional filling \cite{goral2002quantum}. Thus, low entropy samples of dipolar molecules will constitute a new platform for many-body physics and quantum simulation. 

The realization of a Bose-Einstein condensate of dipolar molecules has been elusive for more than two decades, since the first theoretical ideas on their use were explored \cite{demille2002quantum,goral2002quantum}. When the first ultracold gases of dipolar ground state molecules were created in 2008 \cite{ni2008high}, it was found that chemical reactions can lead to relatively fast losses which limited their lifetimes \cite{ospelkaus2010quantum}. In hopes to mitigate such losses, ground state samples of chemically stable species \cite{zuchowski2010reactions} were prepared \cite{takekoshi2014ultracold,molony2014creation,park2015ultracold}, but the problem of short lifetimes remained \cite{ye2018collisions, gregory2019sticky, bause2021collisions}, likely due to loss channels that open up as a result of molecules' complex internal structure and interactions \cite{christianen2019photoinduced}. Such losses prevented efficient evaporative cooling, a workhorse technique in the preparation of atomic quantum gases. Recently, various collisional shielding techniques \cite{cooper2009stable, micheli2010universal, lassabliere2018controlling, karman2018microwave} have led to the production of molecular clouds with reduced losses, but evaporative cooling has remained relatively inefficient \cite{valtolina2020dipolar, schindewolf2022evaporation, bigagli2023collisionally, lin2023microwave}. For fermionic molecules, thanks to their intrinsically lower losses, this has been sufficient to create degenerate Fermi gases \cite{valtolina2020dipolar, schindewolf2022evaporation}. But recent work \cite{bigagli2023collisionally, lin2023microwave} showed that further substantial improvements to collisional shielding are key to reaching quantum degeneracy for bosonic molecules.

\begin{figure*} [t]
  \centering
  \includegraphics[width = 1 \textwidth]{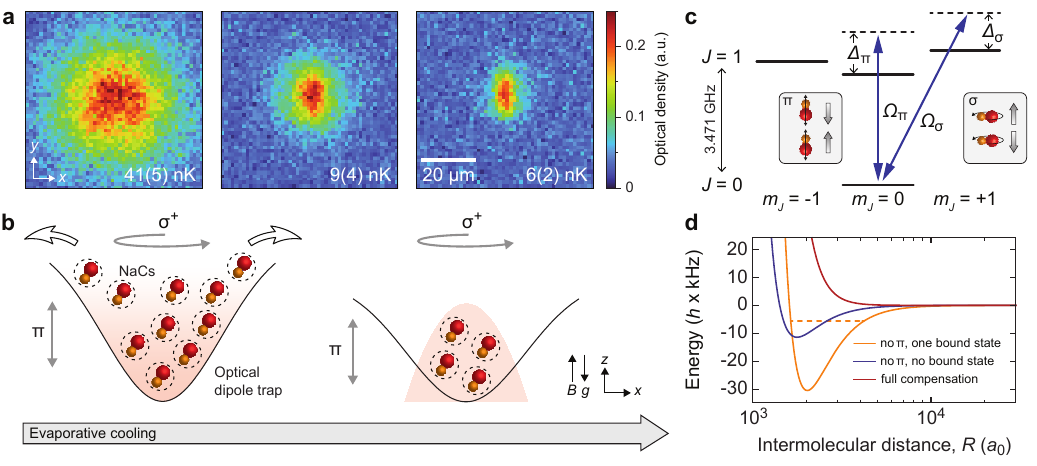}\\
  \caption{\textbf{BEC of dipolar NaCs molecules enabled by microwave shielding.} \textbf{a}, Absorption images of a thermal cloud (left), a partially condensed cloud (middle), and a quasi-pure BEC (right) after 17 ms of time-of-flight expansion. Each image is an average of 20 individual pictures. The clouds, from left to right, correspond to those labeled \textcircled{1}, \textcircled{3}, and \textcircled{5} in Figs. \ref{fig:2} and \ref{fig:3}\textbf{a}. \textbf{b}, Evaporative cooling of NaCs molecules from a thermal cloud to a BEC. The molecules are held in an optical dipole trap and dressed by circularly polarized ($\sigma^+$) and linearly polarized ($\pi$) microwave fields. The collisionally stable molecular gas is cooled by lowering the trap depth forcing out the hottest molecules. Thermal (left) and condensed (right) gases have different density profiles. \textbf{c}, Rotational levels of NaCs, coherently coupled by two microwave fields. The shielded dressed state is a superposition of $|J, m_J\rangle = |0, 0\rangle$, $|1,0\rangle$, and $|1,1\rangle$. The $\sigma^+$ ($\pi$) microwave field has a Rabi frequency $\Omega_{\sigma}$ ($\Omega_{\pi}$) and a detuning $\Delta_{\sigma}$ ($\Delta_{\pi}$) from the $|0,0\rangle \leftrightarrow|1,1\rangle$ ($|0,0\rangle \leftrightarrow|1,0\rangle$) transition. The grey-shaded boxes illustrate the compensation of the $\pi$- and $\sigma^+$-induced dipole moments for a collision of two molecules in the vertical direction: The vertically oscillating dipole moments, induced by the $\pi$ field, interact attractively. Conversely, the rotating dipole moments, induced by the $\sigma^+$ field, interact repulsively. \textbf{d}, Potential energy curves of microwave-shielded molecules approaching in the s-wave channel. The Rabi frequencies and detunings are for the red line $\Omega_{\sigma} = 2\pi\times 7.9$ MHz, $\Delta_{\sigma} = 2\pi\times 8$ MHz and $\Omega_{\pi} = 2\pi\times 6.5$ MHz, $\Delta_{\pi} = 2\pi\times 10$ MHz; for the orange line $\Omega_{\sigma} = 2\pi\times 7.9$ MHz, $\Delta_{\sigma} = 2\pi\times 8$ MHz and no $\pi$ field; and for the blue line $\Omega_{\sigma} = 2\pi\times 7.9$ MHz, $\Delta_{\sigma} = 2\pi\times 17$ MHz and no $\pi$ field. The experimental uncertainty of $\Omega_\sigma$ ($\Omega_\pi$) is 0.3 MHz (0.2 MHz).}
  \label{fig:1}
\end{figure*}

Here, we demonstrate the realization of a Bose-Einstein condensate of dipolar molecules. By strongly suppressing two- and three-body losses, we evaporatively cool ensembles of NaCs molecules from 700(50) nK to 6(2) nK within 3 seconds. Figs.~\ref{fig:1}\textbf{a} and \textbf{b} show sample images during the evaporation sequence and an illustration of the experimental approach, respectively. We reach the critical phase-space density for a BEC with over 2,000 molecules and further evaporate to BECs with 200 molecules and small thermal fractions. The BECs are found to be stable, with a 1/$e$-lifetime of 1.8(1) s. These results show how molecules have achieved a degree of quantum control analogous to that of atoms, drastically expanding the scope of the quantum systems that can be studied.

\section{Microwave Shielding}

For efficient evaporative cooling, collisional losses need to be strongly suppressed. To achieve this, we transfer the molecules into a dressed state using two different microwave fields, one with circular $\sigma^+$-polarization and one with linear $\pi$-polarization. The level diagram for the bare rotational states of NaCs and the microwave frequencies coupling them is shown in Fig.~\ref{fig:1}\textbf{c}.

Microwave shielding as demonstrated so far has a fundamental limit to its effectiveness as there is a trade-off between the suppression of two- and three-body losses \cite{bigagli2023collisionally}. Our current understanding is as follows: When a single circularly polarized microwave field is employed, a superposition of two rotational states induces a rotating dipole moment. At short range, this forms a strong repulsive barrier that prevents two-body loss, with lower losses the stronger the microwave coupling. At long range, dipole-dipole interactions remain attractive (in the s-wave channel). As a result, at an intermolecular distance of $\sim 2,000\, a_0$ an attractive potential well appears (see Fig.~\ref{fig:1}\textbf{d}) which supports field-linked bound states when the microwave coupling is strong~\cite{avdeenkov2003linking, chen2023ultracold}. These bound states can give rise to loss via three-body recombination, hence the more effective the suppression of two-body loss, the stronger the three-body losses. This sets a lower limit to the achievable loss rates and subsequently caps the efficiency of evaporative cooling. The blue and orange curves in Fig.~\ref{fig:1}\textbf{d} show the intermolecular potentials for shielding with a far-detuned and a near-detuned $\sigma^+$ field, limited by two- and three-body losses respectively. 


In order to suppress three-body recombination, field-linked bound states need to be removed while preserving a highly effective suppression of two-body loss. This is achieved by compensating the attractive dipole-dipole interactions at long range while leaving the dipole-induced repulsive barrier at short range unaffected. Dipole-dipole interactions of rotating dipoles induced by a $\sigma^+$ field and oscillating dipoles induced by a $\pi$ field have the peculiar property to be opposite in sign~\cite{karman2018microwave}. By simultaneously dressing the molecules with a circularly- and a linearly-polarized microwave field, we compensate the induced dipole-dipole interactions and minimize the long-range attraction \cite{karman2023upcoming}. This allows the engineering of a purely repulsive intermolecular potential, shown as the red curve in Fig.~\ref{fig:1}\textbf{d}, minimizing both two- and three-body losses. A similar compensation of the induced dipole moment can likely also be achieved by combining microwave and electrostatic fields~\cite{gorshkov2008suppression}.  

\begin{figure*}[t]
  \centering
  \includegraphics[width = \textwidth]{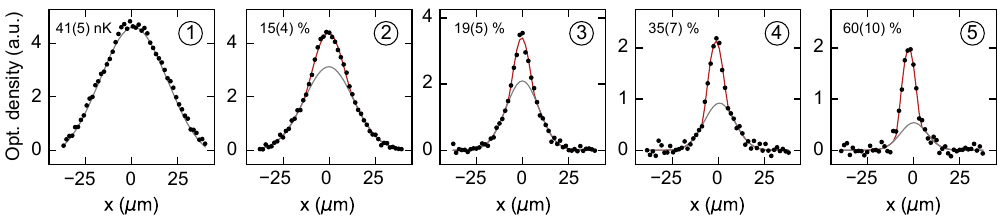}\\
  \caption{\textbf{Formation of the molecular Bose-Einstein condensate.} Each plot shows the 1D profile of absorption images (integrated along the $y$-axis) at different stages of the evaporative cooling sequence, going from a thermal distribution to bimodal distributions to a highly condensed gas (left to right). Red lines show bimodal Thomas-Fermi plus Gaussian fits to the profiles; grey lines separately show the Gaussian part of the fit. On the top-left corner the temperature (for the thermal cloud) and the condensate fraction (for the partially condensed clouds, see Methods) are shown. The temperature is obtained from ballistic expansion. Each plot shows an average of 20 images after 17 ms time-of-flight expansion. The encircled numbers indicate the corresponding data points in Fig.~\ref{fig:3}\textbf{a}.}
  \label{fig:2}
\end{figure*}

In addition to suppressing collisional losses, our microwave dressing scheme helps tuning the molecular interactions to a regime where Bose-Einstein condensation is possible.  In the case of bosonic dipolar molecules, interactions are comprised  of s-wave contact interactions, characterized by the s-wave scattering length, $a_{\rm s}$, and dipole-dipole interactions, characterized by the dipolar length, $a_{\mathrm{dd}} = M d_{\rm eff}^2 / (12 \pi \hbar^2 \epsilon_0)$. Here, $M$ denotes the mass of NaCs, $d_{\rm eff}$ the effective dipole moment, $\hbar$ Planck's constant $h$ divided by $2 \pi$, and $\epsilon_0$ the permittivity of free space. For stable BECs of dipolar particles, the interactions must be repulsive ($a_{\rm s} > 0$), weak ($n_0 a_{\rm s}^3 \ll 1$ and $n_0 a_{\rm dd}^3 \ll 1$, where $n_0$ is the peak number density), and the dipolar interactions should be weaker than contact interactions  ($\epsilon_\mathrm{dd} \equiv a_\mathrm{dd}/a_\mathrm{s} < 1$) \cite{chomaz2022dipolar}. For NaCs molecules dressed by a single $\sigma^+$ field that is sufficiently strong to suppress two-body loss, $a_\mathrm{dd}$ is between about $10,00$ and $25,000\, a_0$, making it hard to fulfill these conditions. With the presence of the $\pi$ field, the dipolar length can be strongly reduced. While in principle the full cancellation of dipolar interactions is possible, in practice we achieve $a_\mathrm{dd} = 1,250\, a_0$, due to finite ellipticity of the $\sigma^+$ field, and $a_{\rm s} \sim 1,500\, a_0$. Full details on the calculation of $a_{\rm s}$ and $a_{\rm dd}$ will be reported in Ref.~\cite{karman2023upcoming}. Conversely, we anticipate that the flexible control of $a_{\rm dd}$ via dressing fields will be a cornerstone to gaining access to strongly interacting dipolar phases with $\epsilon_{\rm dd} > 1$ \cite{buchler2007strongly, schmidt2022self} in future work.

\section{Evaporative Cooling}

Our experiment begins with gases of $30,000$ NaCs molecules in their electronic, vibrational, and rotational ground state, held in a crossed optical dipole trap at a temperature of 700(50)~nK. The molecules are adiabatically prepared in the collisionally-shielded dressed state by sequentially ramping up a circularly-polarized $\sigma^+$ and a linearly-polarized $\pi$ microwave field. We found $\Omega_{\sigma} = 2\pi\times 7.9 (0.3)$ MHz, $\Delta_{\sigma} = 2\pi\times 8$ MHz, $\Omega_{\pi} = 2\pi\times 6.5(0.2)$ MHz, and $\Delta_{\pi} = 2\pi\times 10$ MHz to provide good working conditions for evaporative cooling. Here, $\Omega_\sigma$ ($\Omega_\pi$) and $\Delta_\sigma$ ($\Delta_\pi$) are the Rabi frequency and detuning of the $\sigma^+$ ($\pi$) field. Details on the preparation of the sample and on the generation of the microwave fields are provided in the Methods. 

\begin{figure} [!h]
  \centering
  \includegraphics[width = \columnwidth]{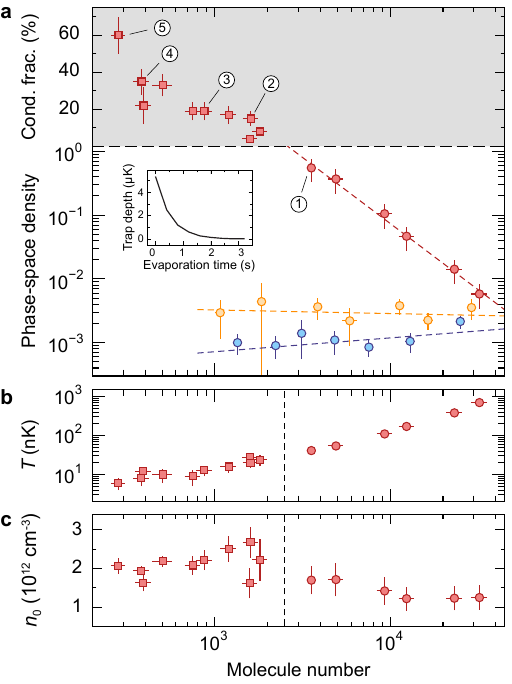}\\
  \caption{\textbf{Evaporative cooling of NaCs molecules to quantum degeneracy.} \textbf{a}, The evolution of the PSD and condensate fraction of the microwave-shielded molecular gas with a purely repulsive intermolecular potential is shown in red. Evaporation attempts of molecular gases with $\sigma^+$-only shielding are shown in orange and blue. The respective microwave parameters are corresponding to the data with the same color in  Fig.~\ref{fig:1}\textbf{d}. The dashed lines are power-law fits to the thermal clouds' PSD used to extract the evaporation efficiency. Error bars show the standard error of the mean of two runs of the experiment. The inset shows the time evolution of the trap depth during the evaporation sequence. \textbf{b}, Decrease of temperature during evaporation. In the thermal regime, temperatures are obtained from time-of-flight expansion of the thermal gas. In the degenerate regime, temperatures are obtained from the expansion of the cloud's thermal component. \textbf{c}, Evolution of peak density during the evaporation. In \textbf{b} and \textbf{c} the vertical dashed lines mark the onset of condensation. For all data points in the figure, circles (squares) denote data for thermal (degenerate) clouds.}
  \label{fig:3}
\end{figure}

We perform forced evaporation by decreasing the depth of the optical dipole trap from $k_\mathrm{B} \times 5.3(0.3)$~$\mu$K to $k_\mathrm{B} \times 40(15)$~nK within 2.8 s, followed by free evaporation for 400 ms by holding the sample in the low-depth trap (see inset of Fig.~\ref{fig:3}\textbf{a}). We record absorption images of the molecular cloud after time-of-flight expansion at various points of the cooling sequence, as shown in Fig.~\ref{fig:2}. Close to the end of the cooling sequence, we start to observe the formation of a BEC through the emergence of a bimodal density distribution. Analyzing the density profiles of the molecular clouds, we find that a bimodal fit captures the shape significantly better than a purely Gaussian fit, as shown in \ref{fig:SI1}. For larger condensate fractions, we observe a marked offset along the $x$-axis between the center of the thermal and condensed components, potentially caused by trap imperfections or repulsion between the two components in time-of-flight. At the end of the cooling sequence, we observe a BEC with a small thermal cloud surrounding it. 

\section{Molecular Bose-Einstein Condensate}

To analyze the cooling process, we determine the phase-space density (PSD), temperature, and peak density of the thermal molecular gas at various points of the cooling sequence (see Fig.~\ref{fig:3}). Their evolution is plotted as a function of molecule number in Fig.~\ref{fig:3}\textbf{a}, \textbf{b}, and \textbf{c}, respectively. Our sample starts at a PSD of $5 \times 10^{-3}$ and the BEC transition is expected at a PSD of 1.202 in a 3D harmonic trap. We reach a PSD of $\sim 1$ at 20~nK with over 2,000 molecules. This aligns with the point where the density profiles show the onset of a bimodal distribution in time-of-flight with a condensed core and a thermal cloud surrounding it. Beyond this point, instead of PSD, we plot the condensate fraction, i.e.~the ratio of the number of molecules in the condensed core and the total number of molecules in the gas, as extracted from bimodal fits. At the end of evaporation, we observe a condensate fraction of $60(10)\%$ for our coldest clouds. From fits to the expansion of the thermal wings, we obtain a temperature of 6(2)~nK. Using the data points in the thermal regime, we also determine the evaporation efficiency to be $d \ln(\rm{PSD}) / d \ln( N) = 2.0(1)$. In the course of the evaporative cooling sequence the PSD increases by more than three orders of magnitude, which far exceeds the gains observed in previous demonstrations of evaporative cooling of molecules~\cite{son2020collisional, valtolina2020dipolar, li2021tuning, schindewolf2022evaporation}.

Interestingly, the peak density of the molecular cloud stays approximately constant during the evaporation, as shown in Fig.~\ref{fig:3}\textbf{c}. In the thermal regime it is around $n_0 = 1.5(3) \times 10^{12}$ cm$^{-3}$, before slightly increasing to $n_0 = 2.0(5) \times 10^{12}$ cm$^{-3}$ in the degenerate regime (see Methods). Compared to atomic BECs, which often reach peak densities above $10^{14}$ cm$^{-3}$, these are unusually low densities, induced by the large value of $a_{\rm s}$. Thanks to these conditions our system is in the weakly interacting regime $n_0 a_{\rm s}^3 \ll 1$ and $n_0 a_{\rm dd}^3 \ll 1$, which ensures that quantum depletion is negligible.

To illustrate the efficacy of the enhanced microwave shielding scheme employed in this work, we also show the evolution of PSD for two cases of $\sigma^+$-only shielding. These two cases correspond to the scattering potentials shown in Fig.~\ref{fig:1}{\bf d}. Following the exact same evaporation ramp of the optical dipole trap as with the enhanced shielding scheme, we now observe an evaporation efficiency of \mbox{-0.2(1)} for the scattering potential with a weaker collisional barrier (blue) and 0.1(1) for the scattering potential with a stronger collisional barrier and a larger long-range attractions (orange). These shielding schemes come with significantly higher loss rates, such that no significant gains in PSD are made for the given sequence. 

\section{Condensate Properties}

\begin{figure}
  \centering
  \includegraphics[width = \columnwidth]{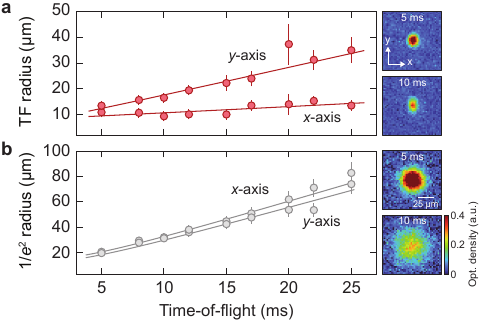}\\
  \caption{\textbf{Time-of-flight expansion.} \textbf{a}, Quasi-pure BEC with a condensate fraction of $>50$~\% released from an elongated trap with trap frequencies $\omega_{\rm x} / (2 \pi) = 23(2)$~Hz and $\omega_{\rm y} / (2 \pi) = 49(3)$~Hz. The solid lines are linear fits to the radii. \textbf{b}, Thermal gas at $T=37(4)$ nK, just before the onset of degeneracy, released from an elongated trap with trap frequencies $\omega_{\rm x} / (2 \pi) = 23(2)$~Hz and $\omega_{\rm y} / (2 \pi) = 53(3)$~Hz. The solid lines are fits to the expansion of the thermal cloud that allow us to retrieve its temperature. For the BEC (thermal gas), data points show the Thomas-Fermi radius (Gaussian $1/e^2$-radius) for both the $x$- and $y$-axis. Each data point is the average of 20 (5) iterations of the experiment.}
  \label{fig:4}
\end{figure}

Having established the phase transition to a BEC, we investigate some of its qualitative properties. First, we observe the expansion dynamics in free flight by releasing a quasi-pure BEC from an elongated trap with aspect ratio $\omega_{\rm x}/\omega_{\rm y} \approx 0.5$. We observe an anisotropic expansion of the condensate and the inversion of aspect ratio, as shown in Fig.~\ref{fig:4}\textbf{a}. This aligns with the expected behavior of matter wave-type expansion of a BEC released from an elongated trap~\cite{mewes1996bose}. It is markedly different from the behavior of a thermal cloud that typically expands isotropically, independent of the trap's aspect ratio. We confirm that a thermal molecular gas just before the onset of condensation expands isotropically when released from a trap with the same aspect ratio, as shown in Fig.~\ref{fig:4}\textbf{b}. In how far dipolar interactions affect the dynamics of the expanding BEC is a question of great interest, which will be addressed in future work. We also note that the observation of an expanding BEC is additional confirmation that a weakly dipolar gas has been created ($\epsilon_\mathrm{dd} < 1$), in contrast, for example, to a self-bound droplet that would not expand in time-of-flight~\cite{chomaz2016quantum}.  

\begin{figure}
  \centering
  \includegraphics[width = \columnwidth]{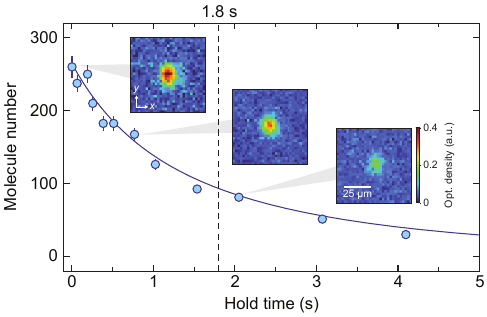}\\
  \caption{\textbf{BEC lifetime.} The BEC is held in the optical dipole trap for a variable hold time and the molecule number is recorded. The initial peak density is $n_0 = 2.0(5) \times 10^{12}$ cm$^{-3}$. Each data point before (after) 1 s is the average of 5 (10) images. The insets are averaged images at different hold times, recorded after 8 ms of time of flight. Errors show the 1$\sigma$ uncertainty from the fits.}
  \label{fig:5}
\end{figure}

Second, we measure the BEC lifetime. As shown in Fig.~\ref{fig:5}, we observe the molecule number in the BEC as a function of hold time in an optical dipole trap with a depth of $k_\mathrm{B} \times 40(15)$ nK. As three-body processes are strongly suppressed thanks to enhanced microwave shielding, we fit the data with a kinetic model that includes one- and two-body losses specific to a BEC (see Methods). We find a 1/$e$ lifetime of the condensate of 1.8~s. From the data we also extract a two-body loss rate of $\beta_{\rm 2B} = 3(1)\times 10^{-13}$ cm$^3$s$^{-1}$, which we believe is mostly due to free evaporation in the relatively shallow trap. This is four orders of magnitude lower than the loss rate of unshielded molecules. 

\section{Conclusion} 
In conclusion, we have created a BEC of dipolar molecules. Leveraging the tunability of dipolar interactions, we have achieved a dramatic suppression of losses and simultaneously created the conditions for a weakly interacting Bose gas. With hundreds of molecules in an identical internal and motional state, the BEC is an ideal starting point for the exploration of strongly dipolar quantum matter. Thanks to its large dipole moment of 4.75 D \cite{dagdigian1972molecular}, NaCs is ideally suited to tune between the weakly and strongly dipolar regime, which is hard to achieve in other dipolar systems, such as magnetic atoms or Rydberg atoms.

With the ability to create stable BECs, dipolar molecules make a significant leap towards becoming a new modality for quantum simulation, quantum information, and the exploration of novel many-body quantum systems. The BEC should give direct access to exotic forms of self-organization in 3D, such as the formation of droplet arrays \cite{kadau2016observing} and macrodroplets \cite{chomaz2016quantum}, predicted for densities and interaction strengths that can be reached from current conditions. In 2D systems, the emergence of strongly interacting superfluids, dipolar crystals, supersolid, and hexatic phases has been predicted \cite{buchler2007strongly, baranov2012condensed}. Furthermore, the BEC should be an ideal starting point to load optical lattices. Stabilized through microwave shielding, the realization of extended Hubbard models with finite tunneling and wide tunability of interactions comes within reach, giving access to Mott insulators with fractional filling \cite{Capogrosso2010}. In particular, it should become possible to realize the long-standing goal of loading optical lattices with unity filling, i.e.~exactly one molecule per lattice site. This will be a critical prerequisite for the realization of spin models with hundreds of interacting spins \cite{gorshkov2011tunable}. Combining the lattice-loaded molecules with microwave dressing schemes, also the formation of topologically-ordered phases \cite{manmana2013topological} or dipolar spin liquids \cite{yao2018quantum, buchler2007three} may come within reach.

\section{Acknowledgements}
We thank Chris Greene, Ahmed Elkamshishy, and Shayamal Singh for helpful discussions and preliminary calculations on the field-linked bound states in the shielding potentials and Rachel Wooten, Tarik Yefsah, and Martin Zwierlein for critical reading and helpful comments on the manuscript. We are grateful to Aden Lam and Claire Warner for important contributions in the construction of the experimental apparatus. We also thank Immanuel Bloch, Tin-Lun Ho, and Vladan Vuleti\'{c} for insightful discussions. We acknowledge Emily Bellingham and Haneul Kwak for experimental assistance. We thank Rohde $\&$ Schwarz for the loan of equipment. This work was supported by an NSF CAREER Award (Award No.~1848466), an ONR DURIP Award (Award No.~N00014-21-1-2721), a grant from the Gordon and Betty Moore Foundation (Award No.~GBMF12340), and a Lenfest Junior Faculty Development Grant from Columbia University. W.Y.\ acknowledges support from the Croucher Foundation. I.S. was supported by the Ernest Kempton Adams Fund. S.W.\ acknowledges additional support from the Alfred P. Sloan Foundation.

%

\newpage

\setcounter{figure}{0}
\makeatletter 
\renewcommand{\thefigure}{Extended Data \@arabic\c@figure}
\makeatother

\newpage

\section{Methods}

{\bf Sample preparation and detection.} The starting point for the experiments in this work are ensembles with about 30,000 ground state NaCs molecules at a temperature of 700(50) nK. At the beginning of forced evaporation, the sample is held in a crossed optical dipole trap with trap frequencies $\omega / (2 \pi) = (45, 78, 162)$~Hz (measured for NaCs ground state molecules). The $x$-dipole trap is elliptical and focused to waists of 108(1)~$\mu$m (horizontal) and 51.5(5)~$\mu$m (vertical); the $y$-dipole trap is almost circular with waists of 117(1)~$\mu$m (horizontal) and 104.5(5) (vertical). Optical trapping light is generated by a 1064 nm narrow-line single-mode Nd:YAG laser (Coherent Mephisto MOPA). At the end of  evaporation the trap frequencies are $\omega / (2 \pi) = (23, 49, 58)$~Hz.

The ensembles of ground state molecules are prepared in three steps. First, overlapping ultracold gases of Na and Cs atoms are created \cite{warner2021overlapping}. Second, weakly bound NaCs Fesh\-bach molecules are assembled via a magnetic field ramp across the Feshbach resonance at $B_\mathrm{res} = 864.1(1)$ G \cite{lam2022high}. The magnetic field points in $z$-direction and sets the quantization axis. Third, NaCs Feshbach molecules are transferred to the electronic, vibrational, and rotational ground state, $X^1\Sigma^+ |v , J \rangle = | 0, 0\rangle$, via stimulated Raman adiabatic passage (STIRAP) \cite{stevenson2023ultracold, warner2023efficient}. The specific hyperfine state of ground state molecules is $|m_{I_{\rm Na}}, \ m_{I_{\rm Cs}}\rangle = |3/2,\ 5/2\rangle$, where $m_{I_{\rm Na}}$ $(m_{I_{\rm Cs}})$ is the projection of the nuclear spin of sodium (cesium) onto the quantization axis. Due to the large $B$ field, the nuclear spin is decoupled from the rotational spin, such that hyperfine substructure does not need to be considered in the microwave shielding process. The STIRAP beams propagate vertically along the $z$-axis, parallel to gravity. In this way, the molecules remain inside the STIRAP beam profiles as they fall and expand in time-of-flight while shielded in the ground state to prevent losses. This is critical for precise thermometry of the molecular gas \cite{bigagli2023collisionally}. 

At the end of time-of-flight expansion, NaCs molecules are detected by ramping down the dressing fields (80~$\mu$s), reversing STIRAP (20~$\mu$s), optical dissociation with a pulse of light (100~$\mu$s) that is resonant with the Cs $6^2 S_{1/2} |F, m_F \rangle = |3, 3\rangle \rightarrow 6^2 P_{3/2} |4, 4\rangle$ transition at high magnetic field, immediately followed by absorption imaging of Cs atoms with a pulse of light (100~$\mu$s) that is resonant with the Cs $6^2 S_{1/2} | 4, 4\rangle \rightarrow 6^2 P_{3/2} | 5, 5\rangle$ transition at high field. Here, $F$ is the total atomic angular momentum and $m_F$ its projection onto the quantization axis. The imaging resolution in our system is 4.5(5) $\mu$m (standard deviation of a Gaussian), comprised of a diffraction-limited resolution of $3$~$\mu$m and momentum diffusion of about $3.5$~$\mu$m of Cs atoms during the 100 $\mu$s imaging light pulse. The resolution is separately confirmed by measuring the the smallest detectable cloud size for a small NaCs BEC.

\begin{figure} 
    \centering
    \includegraphics[width = \columnwidth]{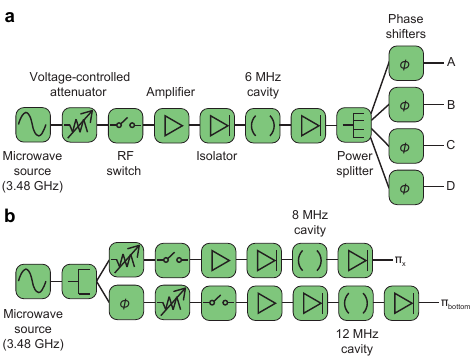}\\
    \caption{\textbf{Microwave setup.} \textbf{a}, Block diagram of the microwave components producing the $\sigma^+$ field.  \textbf{b}, Block diagram of the microwave components producing the $\pi$ field. MHz-level detunings are omitted From the frequencies of the source.}
    \label{fig:SIBlock}
\end{figure}

{\bf Microwave system.} The setup to generate the microwave dressing fields consists of a cloverleaf antenna array producing a circularly polarized ($\sigma^+$) field and two loop antennas, one producing the main linearly polarized ($\pi$) microwave field, and a second one to control the angle between the $\sigma^+$ and $\pi$ fields. The array and the loop antennas are supplied by two separate chains of microwave components, as illustrated in the block diagrams in Fig.~\ref{fig:SIBlock}. The array, described in detail in Ref.~\cite{yuan2023planar}, consists of four resonant loop antennas in a cloverleaf configuration that are fed by a single microwave source. Phase shifts between the loops can be controlled to generate clean circular polarization. The cloverleaf antenna is oriented to emit along the vertical $z$-axis, with circular polarization in the $xy$-plane. The single loop antenna producing the main $\pi$ field is oriented to emit on the horizontal $x$-axis, with linear polarization along the $z$-axis. The second loop antenna, which emits along the $z$-axis, is used to align the polarization vector of the linear $\pi$ field to the circularly polarized $\sigma^+$ field. This is achieved by the careful tuning of the field's amplitude and phase. The molecules are prepared in the microwave-shielded dressed state by first adiabatically increasing the $\sigma^+$ and then the $\pi$ fields. Each intensity ramp is performed within 40~$\mu$s. 

In detail, the chains of microwave components are as follows: The $\sigma^+$ branch starts with a SG12000 signal generator from DS instruments. Its output is fed into a voltage-controlled attenuator (General Microwave, D1954) followed by an RF switch (Mini-Circuits, ZFSWA2R-63DR+). An amplifier (RF Bay, JPA-1000-8000-5) is then used to reach the necessary high power. To reduce the phase noise of the amplifier, responsible for one-body loss of molecules due to transitions to unshielded states, its output is filtered by a 6 MHz-bandwidth cavity (WT Microwave, WT-A10140-Q04). To prevent potential reflections from the cavity that may damage the equipment, we use isolators (Raditek, RADI-3.4-3.6-S3-10WR-10WFWD-H21) on each of its sides. The amplified and filtered signal is then split through a power splitter (Mini-Circuits, ZN4PD1-63-S+) into four different branches, each connected to one of the four antennas of the array. By varying the length of the cable connecting the splitter to each antenna (labelled ``phase shifters'' in the diagram) we realize the relative $90^\circ$ phase shifts that generate $\sigma^+$ polarization. The $\pi$ branch starts with a second SG12000 signal generator. Its output is immediately split into two branches, one with a relative phase shift and attenuation with respect to the other given by a PS6000L phase shifter from DS instruments. The two branches then follow identical paths mirroring the $\sigma^+$ branch, except that the two signals are not split into four but are directly fed into the two loop antennas.

{\bf Rabi frequencies, angle between microwaves and ellipticities.} To measure the $\pi$ Rabi frequency, $\Omega_\pi$, we observe Rabi oscillations between $|0,0\rangle$ and $|1,0\rangle$. The $\sigma^+$ Rabi frequency, $\Omega_\sigma$, is determined via dressed-state spectroscopy (for more details see \cite{zhang2023effects}). Due to the use of narrow-bandwidth cavity filters in the microwave path the direct measurement of the resonant Rabi frequency is not possible. The probe field for the dressed-state spectroscopy is given by the $\sigma^+$ output of the $\pi$ antenna, which does not produce a perfectly pure linear polarization.  To perform dressed-state spectroscopy, we initially increase the $\sigma^+$ field to prepare the molecules in the $|+\rangle$ dressed state, using the detuning reported in the main text. Then we abruptly turn on the field produced by the $\pi$ antenna for a time shorter than a $\pi$-pulse between the $|+\rangle$ and $|-\rangle$ dressed states. By scanning the frequency of the probe field, we find the center frequency of the transition, $\omega$, which in turn gives the $\sigma^+$ Rabi frequency through the relation $\omega - \omega_0 - \Delta_\sigma = \sqrt{\Omega_\sigma^2 + \Delta_\sigma^2}$, where $\omega_0 / 2\pi$ is the known $|0,0\rangle\leftrightarrow |1,1\rangle$ transition frequency and $\Delta_\sigma$ the known detuning from the same transition.

To determine the angle between the fields produced by the $\sigma_+$ and $\pi$ antennas, we combine the knowledge of $\Omega_\pi$ with the measurement of Rabi oscillations between the dressed states $|+\rangle$ and $|-\rangle$. Because the only allowed transitions between these states involve a $\sigma^+$ photon, the Rabi frequency of the dressed state oscillation, $\Omega_{\rm dressed}$, reveals the projection of the $\pi$ field onto the $\sigma^+$ field. Thus, the angle between the vector normal to the $\sigma^+$ field and $\pi$ field is determined by $\arctan\left(\Omega_{\rm dressed} / \left(\Omega_\pi \sin^2(\phi)\right)\right)$. The quantity $\sin^2(\phi)$ accounts for the relative strength of the dressed state sideband transitions \cite{zhang2023effects}, with $\phi$ given by $\sin(2\phi) = 1/\sqrt{1+(\Delta_\sigma/\Omega_\sigma)^2}$. From these measurements we determine that the tilt between the $\pi$ field and the vector normal to the $\sigma^+$ field is less than 1$^\circ$. Since the fields need to be well aligned to ensure the full cancellation of the dipole moment, the size of the tilt angle sets the limit for how well we can compensate the dipole moment \cite{karman2023upcoming}. The realization of a BEC requires relatively weak dipole-dipole interactions, so the tilt angle needs to be small.    

We determine the ellipticity of the $\sigma^+$ field through the direct measurement of the relative Rabi frequencies of the $\sigma^-$ and $\sigma^+$ transitions \cite{yuan2023planar}. We infer the ellipticity of the $\pi$ field from the dressed state spectroscopy. As the projection of the $\pi$ field onto the $\sigma^+$ field is less than 1$^\circ$, the ellipticity of the $\pi$ field must also be less than 1$^\circ$.

{\bf Cloud fitting.} In order to extract the thermal and condensed component of the molecular gas, we perform bimodal fits to the absorption images. To this end, the absorption images are integrated along the $x$- and $y$-direction for better signal-to-noise. The two resulting 1D profiles are simultaneously fitted by the fitting function in direction $\alpha$ ($\alpha = x, y$):
\begin{align*}
    n(\alpha) = & \frac{15}{16} \frac{N_{\rm TF}}{\sigma_{\rm TF,\alpha}}\max\left( 1 - \left(\frac{\alpha - \alpha_{\rm TF}}{\sigma_{\rm TF,\alpha}}\right)^2,0 \right)^{2} \\
    + & \frac{N_{\rm G}}{\sqrt{2\pi}\sigma_{\rm G}} \exp \left( -\frac{(\alpha - \alpha_{\rm G})^2}{2\sigma_{\rm G}^2} \right).
\end{align*}

Along both directions the molecule number in the thermal cloud, $N_{\rm G}$, the molecule number in the condensed cloud, $N_{\rm TF}$, and the Gaussian radius of the thermal cloud, $\sigma_{\rm G}$, are kept identical; $\sigma_{\rm TF,\alpha}$ is the Thomas-Fermi radius, $\alpha_{\rm TF}$ the center of the condensed cloud, and $\alpha_{\rm G}$ the center of the thermal cloud in direction $\alpha$.

Using the results of the fit, the condensate fraction is given by $N_{\rm c} = N_{\rm TF} / (N_{\rm G}+N_{\rm TF}) $ and the peak density by $ n_0 = \left[15^{2/5}/ (8 \pi)\right] \left[ M^3 \bar{\omega}^3 N_{\rm c} / (\hbar^3 a_{\rm s}^{3/2})\right]^{2/5}$ \cite{dalfovo1999theory}, where $\bar{\omega}$ is the geometric mean of the trap frequencies. We note that the calculation of peak density takes into account the effect of s-wave interactions but neglects dipolar interactions. Their inclusion is expected to lead to slightly higher values for the peak densities.

\begin{figure} [t]
  \centering
  \includegraphics[width = \columnwidth]{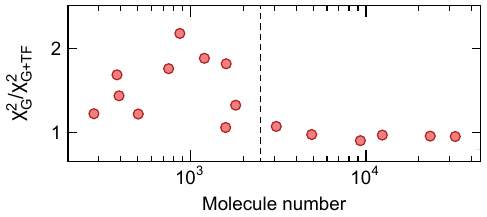}\\
  \caption{\textbf{Comparison of fitting models}. Ratio of  $\chi^2$-values for Gaussian and bimodal fits. The vertical dashed line marks the onset of the phase transition.}
  \label{fig:SI1}
\end{figure}

For absorption images obtained at various points of the evaporation sequence, we compare the quality of the bimodal fits to simple Gaussian fits. We find that in the thermal regime the $\chi^2$ values of the Gaussian and the bimodal fits are nearly identical, while in the degenerate regime the bimodal fit consistently yields a lower $\chi^2$ value. Fig.~\ref{fig:SI1} shows the ratio of the $\chi^2$ values of the Gaussian fits and the bimodal fits, with a behavior similar to that seen in Ref.~\cite{davis1995bose}.    

{\bf Condensate lifetime model.} We extract the two-body loss rate, $\beta_{\rm{2B}}$, of a quasi-pure condensate from a fit of the lifetime data. For the fit, we use the model of Ref.~\cite{soding1999three}, which includes one- and two-body contributions to the loss and takes into account bosonic quantum statistics. 
The differential equation governing the evolution of molecule number as a function of time is
\begin{equation*}
\dot{N} = - \beta_{\rm{2B}} c_2 N^{7/5} -N/\tau_{\rm{1B}},
\end{equation*} 
where $\tau_{\rm{1B}}$ is the one-body lifetime, $c_2 \equiv \left[15^{2/5}/(14\pi)\right]\left[M \bar{\omega}/(\hbar a_{\rm s})\right]^{6/5}$, $M$ is the molecular mass, and $\bar{\omega}$ the geometric mean of the trap frequencies. We use the one-body loss rate $\tau_{\rm{1B}} = 5.0(2)$~s as a fixed parameter in this fit. The one-body loss rate is independently measured by fitting the loss of a molecular cloud at 700~nK for very long hold times ($\sim 20$~s). The one-body loss rate is likely set by off-resonant scattering of the optical trapping light.

\end{document}